\documentclass[conference]{IEEEtran}
\IEEEoverridecommandlockouts
\usepackage{cite}
\usepackage{amsmath,amssymb,amsfonts}
\usepackage{algorithmic}
\usepackage{graphicx}
\usepackage{textcomp}
\usepackage{xcolor}
\usepackage{soul}
\usepackage[a4paper, total={184mm,239mm}]{geometry}
\def\BibTeX{{\rm B\kern-.05em{\sc i\kern-.025em b}\kern-.08em
    T\kern-.1667em\lower.7ex\hbox{E}\kern-.125emX}}

\usepackage{listings}

\definecolor{mybrown}{RGB}{196,104,54}
\definecolor{mygray}{RGB}{72,86,107}
\definecolor{myyellow}{RGB}{236,204,84}

\usepackage[ruled,vlined, linesnumbered]{algorithm2e}
\SetKwRepeat{Do}{do}{while}%
\definecolor{codegreen}{rgb}{0,0.6,0}
\definecolor{codegray}{rgb}{0.5,0.5,0.5}
\definecolor{codepurple}{rgb}{0.58,0,0.82}
\definecolor{backcolour}{rgb}{0.95,0.95,0.92}

\lstdefinestyle{mystyle}{
  backgroundcolor=\color{backcolour},   commentstyle=\color{codegreen},
  keywordstyle=\color{magenta},
  numberstyle=\tiny\color{codegray},
  stringstyle=\color{codepurple},
  basicstyle=\ttfamily\footnotesize,
  breakatwhitespace=false,         
  breaklines=true,                 
  captionpos=b,                    
  keepspaces=true,                 
  numbers=left,                    
  numbersep=5pt,                  
  showspaces=false,                
  showstringspaces=false,
  showtabs=false,                  
  tabsize=2
}

\newtheorem{example}{Example}[section]

\begin{document}

\title{Muzzle the Shuttle: Efficient Compilation for Multi-Trap Trapped-Ion Quantum Computers

\thanks{Accepted in the 2022 Design Automation and Test in Europe (DATE). 

This work is supported by National Science Foundation (NSF) (OIA-$2040667$ and DGE-$2113839$) and seed grants from Penn State Institute for Computational and Data Sciences (ICDS) and Penn State Huck Institute of the Life Sciences.}
}

\author{
\IEEEauthorblockN{Abdullah Ash Saki}
\IEEEauthorblockA{\textit{Dept. of Electrical Engineering} \\
\textit{Pennsylvania State University}\\
University Park, PA \\
ash.saki@live.com}
\and
\IEEEauthorblockN{Rasit Onur Topaloglu}
\IEEEauthorblockA{
\textit{IBM}\\
Poughkeepsie, NY \\
rasit@us.ibm.com}
\and
\IEEEauthorblockN{Swaroop Ghosh}
\IEEEauthorblockA{\textit{Dept. of Electrical Engineering} \\
\textit{Pennsylvania State University}\\
University Park, PA \\
szg212@psu.edu}
}

\maketitle

\begin{abstract}
Trapped-ion systems can have a limited number of ions (qubits) in a single trap. Increasing the qubit count to run meaningful quantum algorithms would require multiple traps where ions need to shuttle between traps to communicate. The existing compiler has several limitations which result in a high number of shuttle operations and degraded fidelity. In this paper, we target this gap and propose compiler optimizations to reduce the number of shuttles. Our technique achieves a maximum reduction of $51.17\%$ in shuttles (average $\approx 33\%$) tested over $125$ circuits. Furthermore, the improved compilation enhances the program fidelity up to $22.68$X with a modest increase in the compilation time.
\end{abstract}


\section{Introduction}
The trapped-ion (TI) quantum bit (qubit) is one of the front-runner technologies to build practical quantum computers. They offer several advantages such as, identical qubits, long coherence times, and all-to-all connectivity among qubits~\cite{ionq}. Several companies such as IonQ and Honeywell are pursuing this technology. Recently, Honeywell reported a trapped-ion system with a high quantum volume (QV) of $1024$~\cite{honeywell-qv-1024}.
Several TI hardware systems~\cite{ionq, honeywell-qv-1024} are already commercially available through quantum cloud services such as Honeywell, IBM Quantum Experience, AWS Braket, and Microsoft Azure. Moreover, TI systems are being used for many practical cases and demonstrations (e.g.,~\cite{honeywell-bmw}).

\begin{figure}
    \centering
    \includegraphics[width=3.2in]{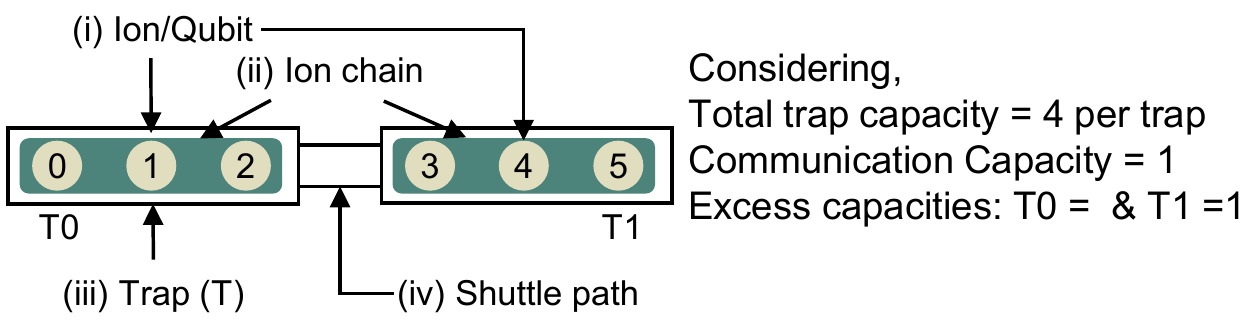}
    \vspace{-3mm}
    \caption{Schematic of a multi-trap TI system. Ions (qubits) (i) are confined inside traps (iii) using DC and oscillatory potentials. Inside a trap, ions form a chain (ii). Traps are interconnected via the shuttle path (iv) which allows movement of ions between traps.\vspace{-3mm}}
    \label{fig:trap-overview}
\end{figure}

The existing TI systems have a smaller number of qubits compared to their superconducting counterparts (IBM's Manhattan quantum processor has $65$ qubits whereas largest known TI system has $11$ qubits~\cite{ionq}). However, roadmaps for larger systems with $50$-$100$ qubits are in place~\cite{honeywell-h5, staq}. Confining many ions in a single trap is a major roadblock for scalable TI systems. With many ions in a single trap, the spacing between ions reduces, making individual ion addressing difficult. Moreover, the gate time becomes slow leading to a longer program execution time. To resolve these issues and to build TI systems with more ions, quantum charge-coupled device (QCCD) based multiple interconnected traps are proposed~\cite{qccd}. Fig.~\ref{fig:trap-overview} shows a schematic of a $2$-trap system interconnected by a \emph{shuttle path}. 

In a multi-trap system, sometimes computation is required on data from ions situated in different traps. In such cases, one ion needs to be shuttled (moved) from one trap to the other so that the ions are co-located, and the gate operation can be performed. A compiler adds shuttle operations to a quantum program to satisfy the inter-trap communication. The shuttle operation is expensive as it degrades quantum gate fidelity. Therefore, minimizing the number of shuttle operations is beneficial and desired. Murali et al.~\cite{murali-ti} performed extensive architectural studies for multi-trap trapped ion systems. They developed a QCCD compiler to generate hardware executable programs from high-level versions and a simulator~\footnote{Ref.~\cite{qccd-github} is the accompanying code-base for~\cite{murali-ti}.} with experimentally calibrated values.
Although the first attempt to build a compiler for multi-trap TI systems, the compiler in~\cite{murali-ti} suffers from several inefficiencies (described in detail in Section~\ref{sec:qccd-compiler}) which lead to a higher number of shuttle operations. This, in turn, inevitably increases ion-chain energy and degrades program fidelity.

In this paper, we optimize the compiler to reduce the number of shuttles and improve the program fidelity in the process. We make the following contributions in this paper:\\
    $\bullet$ We introduce three optimization heuristics, i.e., future ops-based shuttle direction policy, opportunistic gate re-ordering, and nearest-neighbor-first re-balancing (with better ion selection).\\
    $\bullet$ We evaluate our proposals across $5$ NISQ and $120$ random quantum benchmarks and compare with~\cite{murali-ti} to showcase the efficiency of our optimizations.\\
    $\bullet$ We report the improvement in fidelity with minor compilation time overhead.

The rest of the paper is organized as follows: Section~\ref{sec:basics} describes the basics of trapped-ion systems. Section~\ref{sec:qccd-compiler} explains the limitations of the existing compiler and presents our algorithms for compiler optimizations.
Section~\ref{sec:evaluation} reports numerical values of the number of shuttles, fidelity, and compilation time. Finally, the conclusion is drawn in Section~\ref{sec:conclusion}.

\section{Basics}\label{sec:basics}
In this section, we discuss the basics of trapped-ion quantum computers and terminologies used in the paper. 

\begin{figure}
    \centering
    \includegraphics[width=3.2in]{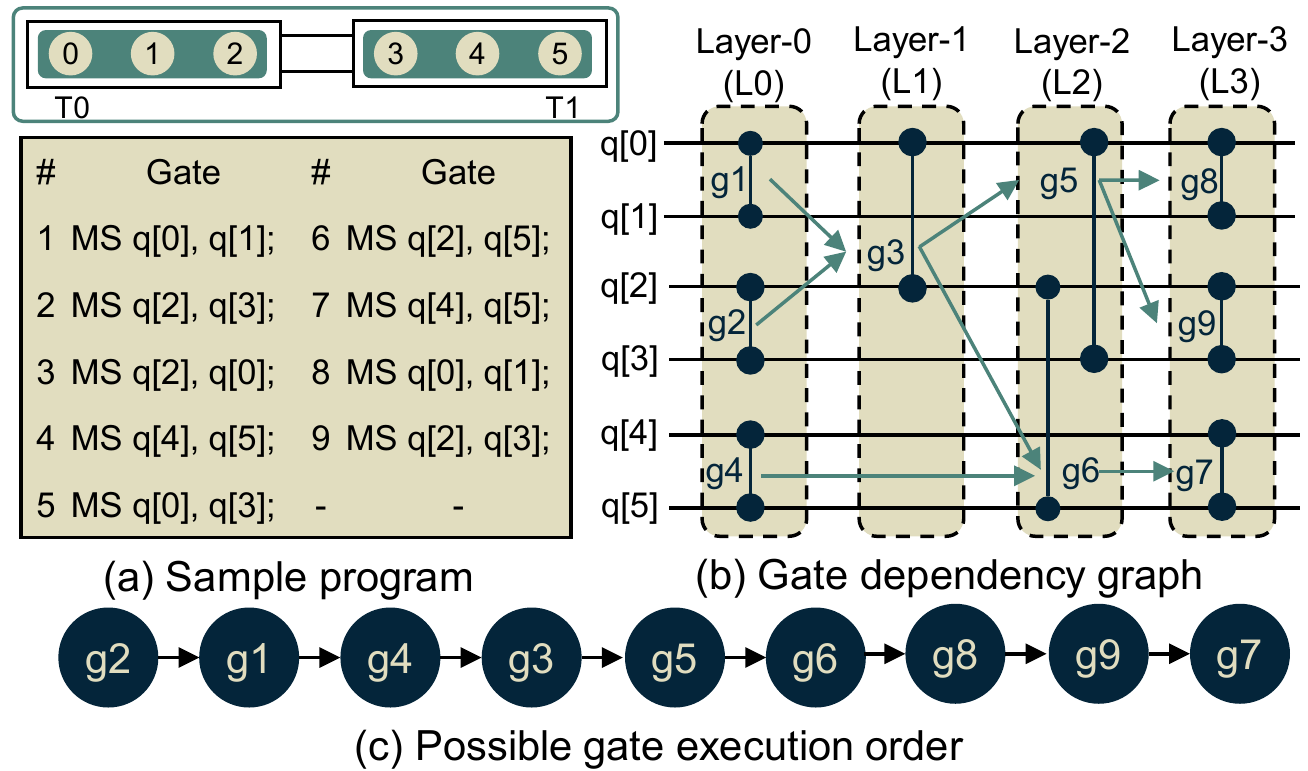}
    \caption{(a) A sample quantum program consisting of 2-qubit gates. (b) Gate dependency graph of the sample program. (c) A possible gate order that satisfies the dependency graph. (Inset on top left: Possible allocation of $6$ ions in the $2$-trap system).\vspace{-3mm}}
    \label{fig:layers-and-order}
\end{figure}

\subsection{Gate dependency graph}
A quantum program is a sequence of quantum gates. Fig.~\ref{fig:layers-and-order}a shows a sample quantum program consisting of 2-qubit MS gates. A quantum program can be converted to a gate dependency graph (a directed acyclic graph, DAG) which consists of \emph{layers}. Gates in a layer are independent of each other but depends on one or more gates from previous layers. Fig.~\ref{fig:layers-and-order}b shows the dependency graph for the sample program in Fig.~\ref{fig:layers-and-order}a. Gates $g5$ and $g6$ are independent of each other (both in Layer--$2$ or L$2$) as they work on different sets of qubits. However, both $g5$ and $g6$ depends on $g3$ which means $g5$ or $g6$ cannot be executed before $g3$. Fig.~\ref{fig:layers-and-order}c shows a possible gate order that satisfies the DAG in Fig.~\ref{fig:layers-and-order}b.

\subsection{Trapped-ion Quantum Computer}

\subsubsection{Trap details}
A trapped-ion system consists of micro-fabricated surface electrode traps which confine ions like Yb or Ca using electromagnetic fields~\cite{ionq}. 
We schematically show different components of a trapped-ion system in Fig.~\ref{fig:trap-overview}. 
A single trap can accommodate a fixed number of ions. We name this \emph{total trap capacity}. During the initial allocation of ions, a part of the total trap capacity is loaded with ions and the remaining capacity (termed as \emph{communication capacity}) is kept unoccupied to allow for shuttled ions from other traps. 
During program execution, ions may need to move (get shuttled) between traps which will free-up one trap and fill-up another trap. The number of free-spaces is termed as \emph{excess capacity} (EC) and defined as ``total trap capacity $-$ number of ions in a trap''. Inside a trap, gates are executed serially (technology constraint) while different traps can have parallel gates.

\begin{figure}
    \centering
    \includegraphics[width=2.5in]{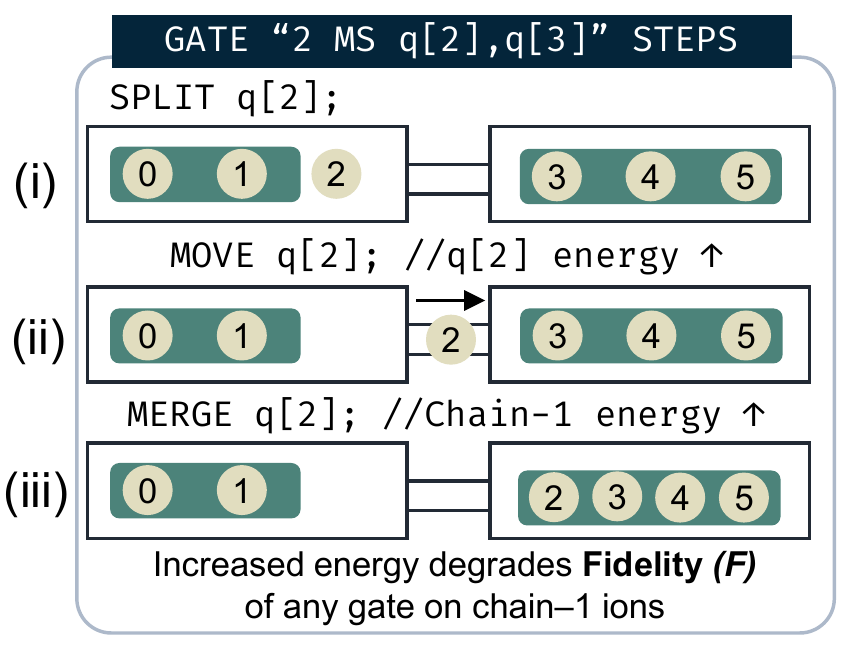}
     \vspace{-3mm}
     \caption{Shuttle steps to bring ions $2$ and $3$ in the same trap. \vspace{-6mm}}
    \label{fig:Shuttle steps}
\end{figure}
\subsubsection{Need for a shuttle operation}
Qubits (ions) are all accessible directly inside a trap in trapped-ion systems. 

For example, the $1^{st}$ gate \texttt{MS q[0], q[1]} in the sample program (Fig.~\ref{fig:layers-and-order}a) can be directly executed as both ions $1$ and $2$ are located in the same trap T$0$. However, the $2^{nd}$ gate in the sample program, \texttt{MS q[2], q[3]}, cannot be directly executed as ion $2$ is in T$0$ and ion $3$ is in T$1$. 
One of the ions need to be shuttled to bring both ions into the same trap.

\subsubsection{Gate fidelity model}
Quantum gates in existing quantum computers including TI systems are erroneous. Ref~\cite{murali-ti} presents an analytical gate fidelity model for TI systems: Fidelity $F = 1 - \Gamma \tau - A(2\bar{n} + 1)$. 
Here, $\Gamma = $ trap heating rate, $\tau =$ gate time, $\bar{n} =$ motional mode or vibrational energy of an ion-chain, $A$ is a scaling factor that varies as $\# qubits/log(\# qubits)$. The simulator in~\cite{murali-ti} uses experimental values of $\Gamma$, $\tau$, $A$, and $\bar{n}$~\cite{fm-gate, shuttle-time}. The values are reported in the paper~\cite{murali-ti} and are embedded in the GitHub code-base~\cite{qccd-github}. We omit those values in this paper for brevity.

\subsubsection{Shuttle steps and impact on fidelity}
The shuttle operation involves several steps as depicted in Fig.~\ref{fig:Shuttle steps}. First, ion $2$ is split from the ion-chain, and then, moved from T$0$ to T$1$. The movement adds energy to the ion. Then, ion $2$ is merged with the other chain, and gate \texttt{MS q[2],q[3]} can be executed. This merge operation increases the vibrational energy ($\bar{n}$) of the chain in T$1$. As motional mode $\bar{n}$ is now higher, the subsequent gate operations in this chain will experience a lower gate fidelity ($F$). 

\section{Compiler Optimizations}\label{sec:qccd-compiler}
In this section, we describe our compiler optimization algorithms. We also discuss the key policies of the QCCD compiler~\cite{murali-ti} along with its limitations and show how our approaches can address them.

\subsection{Shuttle direction policy}\label{subsec:lookahead}
Shuttle direction policy dictates which ion will be moved to execute a $2$-qubit gate. The shuttle direction policy in~\cite{murali-ti} uses excess capacity as the deciding factor. The policy is illustrated in Listing~\ref{shuttle-policy}. 
 
\begin{lstlisting}[language=Python, caption=Shuttle direction policy~\cite{murali-ti, qccd-github}., label=shuttle-policy]
if excess_cap0 < excess_cap1:
    Move Trap0 --> Trap1
elif excess_cap0 == excess_cap1:
    Move 1st ion of the gate
else:
    Move Trap1 --> Trap0
\end{lstlisting}

\subsubsection{\textbf{Issue with excess capacity-based policy}}
\begin{figure}
    \centering
    \includegraphics[width=3in]{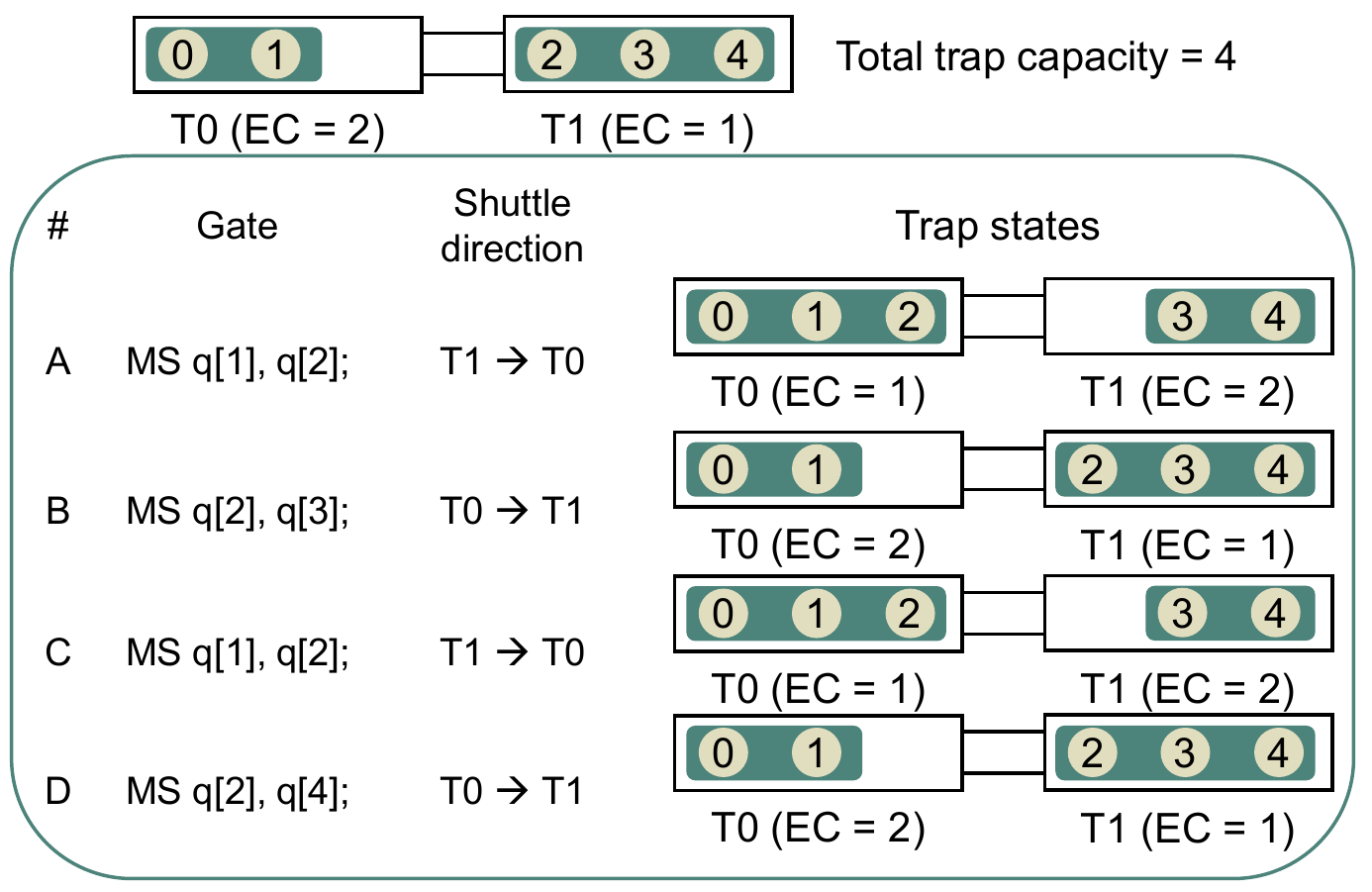}
    \vspace{-3mm}
    \caption{Issues with the shuttle direction policy used in~\cite{qccd-github, murali-ti}. Excess capacity-based logic can lead to repeated shuttles.\vspace{-3mm}}
    \label{fig:shuttle-policy-issue}
\end{figure}
The excess capacity based shuttle direction policy can result in repeated shuttle between traps. We illustrate the issue with an example in Fig.~\ref{fig:shuttle-policy-issue}. Consider a $2$-trap system with total trap capacity of $4$ ions per trap. Consider, there are $2$ ions in T$0$ and and $3$ ions in T$1$. Therefore the excess capacities (ECs) of the traps are EC(T$0$) = $4 - 2 = 2$ and EC(T$1$) = $4 - 3 = 1$. 

Next, consider $4$ gates to be executed starting with \texttt{MS q[1],q[2]}. 
As EC(T$0$) $>$ EC(T$1$), according to the shuttle policy in~\cite{murali-ti} (Listing~\ref{shuttle-policy}), ion $2$ will be moved from T$1$ to T$0$. The trap states are updated after this shuttle, and new excess capacities are EC(T$0$) $= 1$ and EC(T$1$) $= 2$. For the next gate \texttt{MS q[2],q[3]}, ion $2$ will again be moved back to T$1$ according to the shuttle policy. Likewise, the next $2$ gates will also require shuttles. Therefore, $4$ shuttles are required and ion $2$ is shuttled back-and-forth between T$0$ and T$1$.

\subsubsection{\textbf{Proposed future ops-based shuttle direction}}
We propose a future operations-based policy. Suppose, $ion_A$ belongs to $trap_A$, and $ion_B$ belongs to $trap_B$. To implement gate($ion_A$, $ion_B$), one of the ions needs to be shuttled. To make the decision, our heuristics algorithm computes \emph{move score} for each ion as defined below:

\begin{itemize}
    \item $ion_{A(A \rightarrow B)}$ move score $=$ \# $ion_A$ gates in $trap_B$ $+$ \# $ion_B$ gates in $trap_B$
    \item $ion_{B(B \rightarrow A)}$ move score $=$ \# $ion_A$ gates in $trap_A$ $+$ \# $ion_B$ gates in $trap_A$
\end{itemize}

An $ion_{A(A \rightarrow B)}$ move score greater than $ion_{B(B \rightarrow A)}$ move score means that keeping both ions in $trap_B$ will satisfy more future gates than if both ions are kept in $trap_A$. Therefore, moving $ion_A$ to $trap_B$ will be more beneficial and vice-versa. 

Consider the $4$ gate-program in Fig.~\ref{fig:shuttle-policy-issue}. Suppose, $ion_A = 1$, $ion_B = 2$, $trap_A =$ T$0$, and $trap_B =$ T$1$. To perform Gate-A, our logic will look-up the remaining $3$ gates and compute the move score as tabulated in Table~\ref{tab:move-score}. As $ion_{A(A \rightarrow B)}$ move score $>$ $ion_{B(B \rightarrow A)}$ move score, $ion_A = 1$ will move from $trap_A$ (T$0$) to $trap_B$ (T$1$) for Gate-A according to our optimized logic. Moving ion $1$ will satisfy requirement for the remaining $3$ gates in this case, and thus, we will require only $1$ shuttle compared to $4$ as in the previous case.

\begin{table}[]
    \centering
    \caption{Move Score Computation Example}
\vspace{-1mm}    
\begin{tabular}{ccc}
    \# $ion_A$ gates in $trap_B$     &  1 & (Gate-C)\\
    \# $ion_B$ gates in $trap_B$     &  2 & (Gate-B and D)\\
    \hline
    $ion_{A(A \rightarrow B)}$ move score & 3 & -\\
    & \\
    \# $ion_A$ gates in $trap_A$ & 0 & -\\
    \# $ion_B$ gates in $trap_A$ & 1 & Gate-C\\
    \hline
    $ion_{B(B \rightarrow A)}$ move score & 1 & -
    \end{tabular}
\vspace{-3mm}    
    \label{tab:move-score}
\end{table}

\begin{figure}
    \centering
    \includegraphics[width=2.8in]{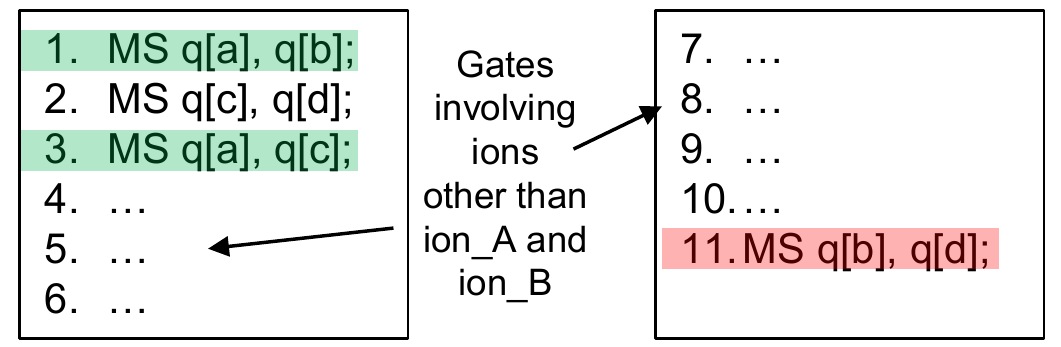}
    \vspace{-3mm}
    \caption{Gate proximity consideration. Distant gates are excluded.\vspace{-4mm}}
    \label{fig:priority}
\end{figure}

\subsubsection{\textbf{Gate proximity consideration}}
While computing the number of future gates, we adopt a proximity-based approach. We argue that not all gates involving $ion_A$ and $ion_B$ have same priorities in the decision making. If there are many other gates between two gates involving $ion_A$ and/or $ion_B$, then we flag the later gates as distant (low proximity) and exclude them from \emph{move score} computation. The proximity between two gates involving $ion_A$ and $ion_B$ is a design parameter in our compiler optimization. The distance should not be too low as it should exclude most future gates from consideration and should not be too high as distant future gates may not represent ion locations correctly. From our analysis, setting the proximity parameter to $6$ provides good results. 

We illustrate the proximity-based approach in Fig.~\ref{fig:priority}. Gate $1$ and $3$ involves either $1$ or both of $ion_A$ and $ion_B$. There is one other gate between these two gates, and thus, distance $=$ $1$. As distance $1 < 6$, gate $3$ is considered. The next gate involving $ion_A$ and/or $ion_B$ is gate $11$. The distance between gate $3$ and gate $11$ is $7$ which is greater than set threshold of $6$.
Thus, gate $11$ (and any related gate after gate $11$) is marked as \emph{low-proximity} and excluded from score computation.

\subsubsection{\textbf{Comment on the complexity}}\label{subsubsec:complexity-shuttle-dir}
The future ops-based shuttle policy checks remaining  gates for each gate requiring shuttle. It can lead to $O(n^2)$ time complexity. With gate proximity consideration, the complexity becomes $O(n.k)$ where $k \leq n$. Moreover, not all gates require shuttles and do not invoke the shuttle direction related computations. Combined with gate proximity consideration, this keeps the value of $n$ and the compilation time in check even for very large circuits.

\subsection{Gate execution order}

\begin{algorithm}[htbp]
\SetAlgoLined
\KwIn{active gate, old\_destination, gate dependency graph, gate order, remaining gates}
\KwOut{new gate order}

active layer $\leftarrow$ the $layer$ the $active~gate$ belongs to\;
check\_gates $\leftarrow$ empty\;
\For{layer $\in$ $1^{st}$ layer to active layer}{
    \For{gate $\in$ layer}{
        \If{gate $\in$ remaining gates and gate $\neq$ active gate}{
            check\_gates.append(gate)\;
        }
    }
}
\For{gate $\in$ check\_gates}{
    Find source trap for the \emph{gate} using future-ops shuttle policy\;
    \If{source\_trap == old\_destination}{
        new\_gate\_order $\leftarrow$ Remove $gate$ from gate order and insert before active gate;
        break\;
    }
}
\caption{Re-order gate execution}
\label{algo:reorder}
\end{algorithm}

For gate execution order, we keep the baseline earliest-ready-gate-first heuristics as in~\cite{murali-ti}. It finds the order by topologically sorting the gate dependency graph. In some cases, the favorable shuttle direction computed by the future ops-based shuttle direction policy may not be achievable if the destination trap is full. In such cases, we reorder the gate execution sequence to free-up the trap maintaining the gate dependency graph. We name the gate to be executed as \emph{active gate}. First, we identify the \emph{layer number} of the active gate (e.g., if $g8$ from Fig.~\ref{fig:layers-and-order}b is the active gate, then layer number is $3$). Any pending gate in this layer and preceding layers are candidates as they can be executed without breaking the gate dependency graph. The algorithm checks each candidate gate. If a candidate gate can free-up a space in the destination trap, then it is moved up the gate execution order before the active gate, and it becomes the new active gate. The algorithm is detailed in Algorithm~\ref{algo:reorder}.

\begin{figure}
    \centering
    \includegraphics[width=3.4in]{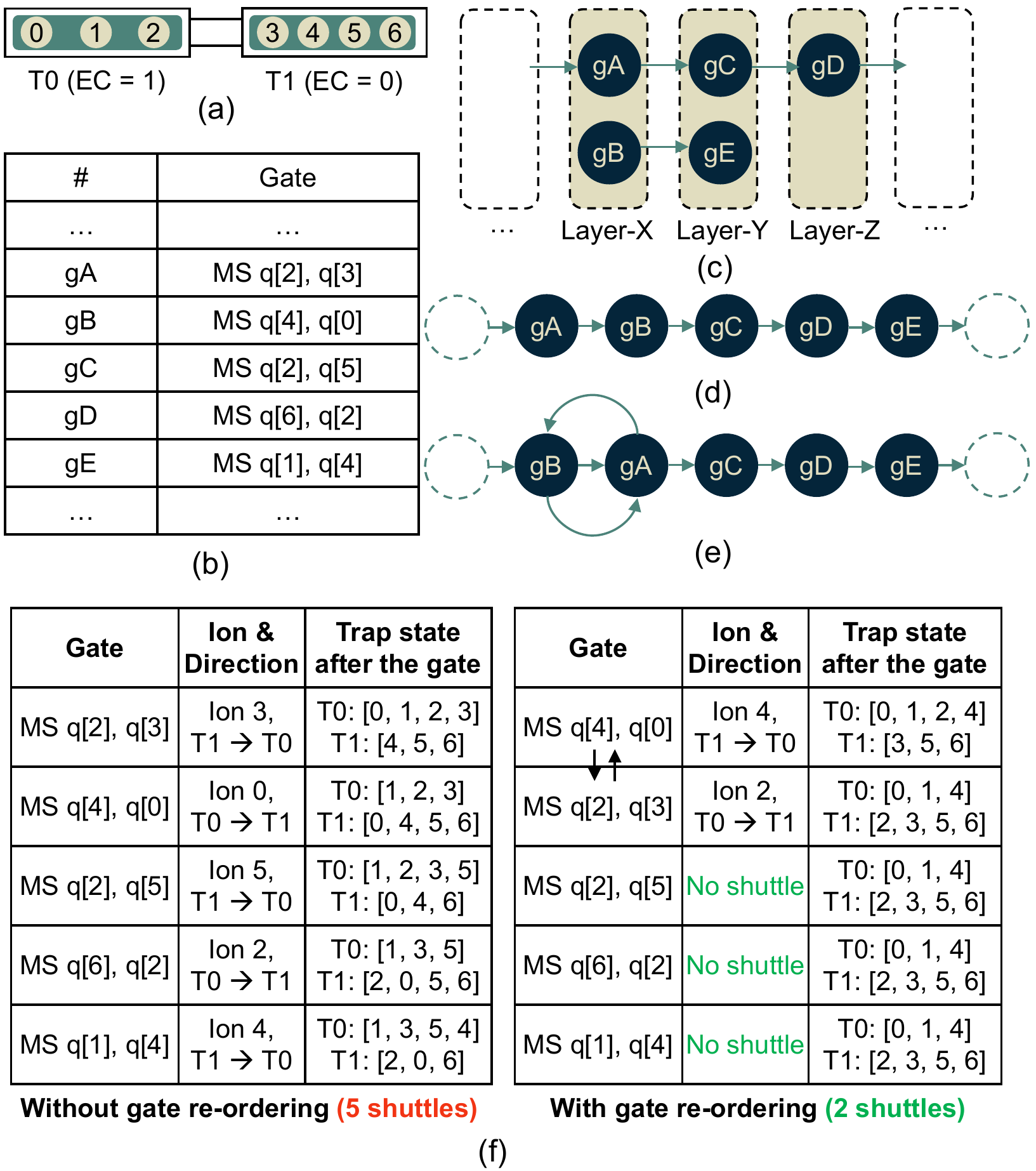}
    \vspace{-3mm}    
    \caption{An example of opportunistic gate re-ordering. (a) An example trap state. (b) Partial quantum program. (c) Gate dependency graph of the partial program. (d) Baseline gate execution order. (e) Re-ordered gate sequence to free-up T$1$. (f) Illustration of shuttle reduction with gate re-ordering.}
    \vspace{-3mm}    
    \label{fig:gate-reorder-example}
\end{figure}

\begin{example}
Fig.~\ref{fig:gate-reorder-example} illustrates an example of gate re-ordering. Gate-A (gA) requires a shuttle, and the favorable direction is moving ion $2$ from T$0$ to T$1$ (as ion $2$ has more operations in T$1$). However, with present trap state (Fig.~\ref{fig:gate-reorder-example}a) T$1$ is full and cannot accept an incoming ion. For such a case, the gate re-ordering logic will be invoked. The logic checks candidate gates (pending gates in the active-layer, Layer-X in this example, and preceding layers). Gate-B in Layer-X is a candidate gate. This gate (gB) also requires a shuttle, and the direction is moving ion $4$ from T$1$ to T$0$ (as ion $4$ has more operations in T$0$). Thus, the re-ordering the baseline order in Fig.~\ref{fig:gate-reorder-example}d to the order in Fig.~\ref{fig:gate-reorder-example}e (i.e., executing gB before gA) will free up a space in T$1$. Next, gA can execute with its favorable direction (ion $2$, T$0$ $\rightarrow$ T$1$). Fig.~\ref{fig:gate-reorder-example}f illustrates how re-ordering gates to support favorable shuttle direction can save shuttles in the example partial program.
\end{example}

\subsubsection{\textbf{Comment on the complexity}}\label{subsubsec:complexity-reorder}
The algorithm checks every \emph{pending gate} in the active layer and the preceding layer(s), and for each pending gate computes a shuttle direction. Again, it has a time-complexity of $O(n^2)$. However, the number of pending gates is typically small even for large circuits. This keeps compilation time in check (supported by numerical results).

\subsection{Resolving traffic blocks}
The compiler~\cite{murali-ti} incorporates logic to resolve traffic blocks in a shuttle path. If a trap is full, it cannot receive any ion which potentially creates a traffic block. The compiler uses a minimum-cost-maximum-flow (MCMF) algorithm to move an ion from a full trap and resolves the traffic block.
In~\cite{murali-ti}, the re-balancing logic always starts searching for a destination trap from $trap-0$. This may result in an inefficient re-balancing which is illustrated with the following example.

\begin{figure}
    \centering
    \includegraphics[width=3.2in]{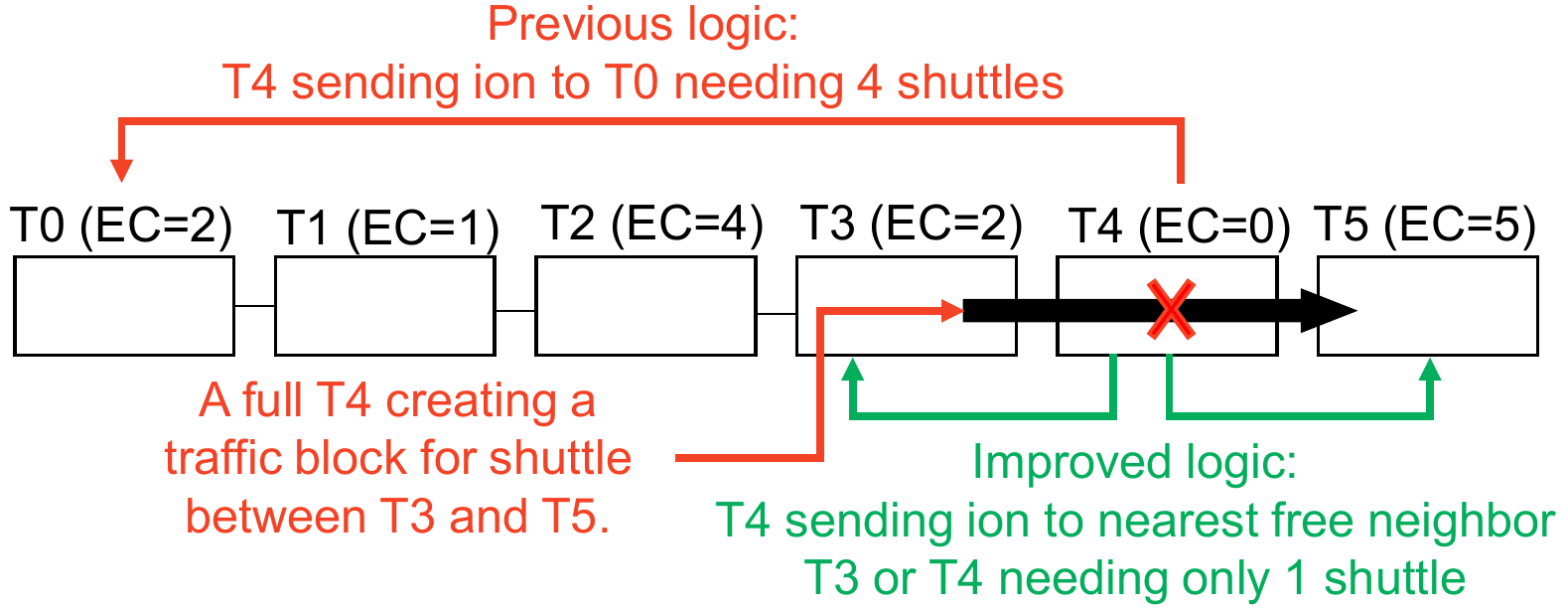}
    \caption{Trap re-balancing logic for traffic block resolution: the problem with the previous logic and a fix with our improved logic. \vspace{-8mm}}
    \label{fig:traffic-block}
\end{figure}

\subsubsection{\textbf{Issue with re-balancing logic}}
Consider, an ion needs to shuttle between T$3$ and T$5$ (Fig.~\ref{fig:traffic-block}). However, T$4$ is full creating a traffic block. 
To remove the traffic block, an ion from T$4$ needs to be moved to another trap. Therefore, we need another trap (destination) with excess capacity ($> 0$). With the present logic in QCCD-simulator~\cite{murali-ti}, the search for a destination trap always starts with T$0$. In the example in Fig.~\ref{fig:traffic-block}, T$0$ has excess capacity, therefore T$4$ will send an ion to T$0$ to re-balance the trap (i.e., free-up traffic block). This will require $4$ shuttles. However, the neighboring traps of T$4$ (T$3$ and T$5$) also have excess capacities. Thus, moving to either T$3$ or T$5$ would require only $1$ shuttle.

\begin{algorithm}
\SetAlgoLined
\KwIn{blocking\_trap, all\_traps, trap\_topology}
\KwOut{destination\_trap}

source\_trap $\leftarrow$ blocking\_trap\;
candidate\_dest\_dist $\leftarrow$ empty hash table\;
\For{candidate\_dest\_trap $\in$ all\_traps}{
    \If{candidate\_dest\_trap $\neq$ source\_trap $and$ trap.excess\_capacity $> 0$}{
        distance $\leftarrow$ shortest distance between source\_trap and candidate\_dest\_trap on trap\_topology\;
        candidate\_dest\_dist[candidate\_dest\_trap] = dist\;
    }
}
destination\_trap $\leftarrow$ key (candidate\_dest\_trap) with the smallest distance in candidate\_dest\_dist\;
\caption{Nearest-neighbor-first re-balancing}
\label{algo:nnf}
\end{algorithm}

\subsubsection{\textbf{Nearest-neighbor-first re-balancing logic}}
We improve the re-balancing logic by searching from nearest-neighbor traps first. The algorithm (Algorithm~\ref{algo:nnf}) first filters out traps with $0$ excess capacities and creates a list of candidate (destination) traps that can accept an ion. 
Finally, the nearest candidate trap is selected as the destination.

\textbf{Max-score shuttle ion selection:}
Besides selecting the destination trap intelligently, we add another optimization in the logic to select a better ion to move. We apply the following heuristics: the ion should have a high number of gates in the destination trap and a low number gates in the source trap. We compute a \emph{score} = ($wd$ $\times$ \# gates in destination $-$ $ws$ $\times$ \# gates in source) for each ion in the source trap.
Usually, we set $wd = ws = 0.5$. If \# gates in destination and \# gates in source are equal we make $wd = 0.49$ and $ws = 0.51$ to avoid score being $0$. 
Finally, ion with highest score is moved.

\subsubsection{\textbf{Comment on the complexity}}\label{subsubsec:complexity-nnf}
Finding a neighboring trap with free spaces has a linear time complexity. As the number of traps are small, searching for a candidate trap is fast. For the \emph{max-score shuttle ion} algorithm, the number of ions to consider becomes a constant as we have a fixed source trap. Thus, it also has a  complexity of $O(constant \times n )$. In~\cite{murali-ti}, the authors show that $15-25$ ions per trap is suitable NISQ applications which bounds the $constant$ to $25$.

\section{Evaluation and Discussions}\label{sec:evaluation}
\begin{table}[b]
\centering
\caption{Reduction in the number of shuttles}
\label{tab:shuttle-reduction}
\begin{minipage}{8cm}
\centering
\def\arraystretch{1.5}\tabcolsep 2pt
\def\thefootnote{a}\footnotesize

\begin{tabular}{ccccccc}
\hline
Benchmark & Qubits & 2Q gates & \cite{murali-ti} & This Work & $\Delta (\downarrow)$ & \%$\Delta$\\
\hline
Supremacy   & 64 & 560  & 365   & 223   & 142   & 38.90\%\\
QAOA        & 64 & 1260 & 1552  & 957   & 595   & 38.34\%\\
SquareRoot  & 78 & 1028 & 717   & 355   & 372   & 51.17\%\\
QFT         & 64 & 4032 & 241   & 196   & 45    & 18.67\%\\
QuadraticForm       & 64 & 3400  & 228    & 164    & 64    & 28.07\%\\
Random          & 60-75 & 1438 (413)   & 1048     & 775 (270)     & 273 (109)     & 26\% (6)\\

\hline
\end{tabular}
\end{minipage}
\end{table}

\subsection{Evaluation Setup}\label{subsec:eval-setup}

\textbf{Experimental Platform:}
All simulations are run on an Ubuntu $20.04$ virtual machine with $8$ GB RAM on a Windows $10$ host with Intel i$7$-$9700$k $3.60$ GHz.

\textbf{Benchmarks:}
To showcase the efficacy of the compiler optimizations, we choose several NISQ benchmarks from~\cite{murali-ti} and Qiskit circuit library. The benchmark suite includes circuit from Google's supremacy experiment, quantum approximate optimization algorithm (QAOA), quantum Fourier transform (QFT), Square Root, and QuadraticForm~\cite{quadratic-form} (quadratic form finds it application in constrained polynomial binary optimization problems). Besides the NISQ benchmarks, we test our compiler with $120$ random circuits. The random circuits are of sizes $60$, $65$, $70$, and $75$ qubits. For each size, we take $30$ randomly generated circuits with average $1438$ 2-qubits gates ($\sigma \approx 413$). For random circuits, we tabulate the mean value with standard deviation in parentheses for performance metrics. 

\textbf{Hardware model:}
For a fair comparison, we use the same hardware model as in~\cite{murali-ti}. We consider the ``L6'' trap topology as in~\cite{murali-ti} where $6$ traps are connected in a linear fashion (Fig.~\ref{fig:traffic-block}). Each trap has a total capacity of $17$ with a communication capacity of $2$ per trap.
To get the program fidelity estimates, we leverage the QCCD simulator~\cite{murali-ti} which includes experimental operation time and gate fidelity models.

\subsection{Number of shuttles}
Table~\ref{tab:shuttle-reduction} shows the reduction in number of shuttle operations for this work compared to~\cite{murali-ti} for benchmarks listed in Section~\ref{subsec:eval-setup}. We observe a $\approx  19\%$ to $51\%$ reduction in shuttles. Our compiler outperforms (results in significantly less number of shuttles) the QCCD compiler in~\cite{murali-ti} for every circuit in the test suite of $125$ circuits. This empirically supports the stability of our algorithms.

The $5$ NISQ benchmarks have structured but different 2-qubit gate pattern~\cite{murali-ti}. For example, Supremacy and QAOA circuits have \emph{nearest neighbor gate} pattern, and for both of them we observe $\approx 38\%$ reduction. The QFT and the QuadraticForm circuits have \emph{all-to-all} connectivities. For such circuits, moving one ion satisfies many future gates, and thus, they tend to have a smaller number of shuttles. The SquareRoot circuit has \emph{short and long-range} gates, and results indicate that we may get best reductions for such patterns. The random circuits have a more unstructured pattern. Our compiler works for random gate pattern as well achieving $26\%$ reduction on average.

\subsection{Program fidelity improvement}
\begin{figure}
    \centering
    \includegraphics[width=3.0in]{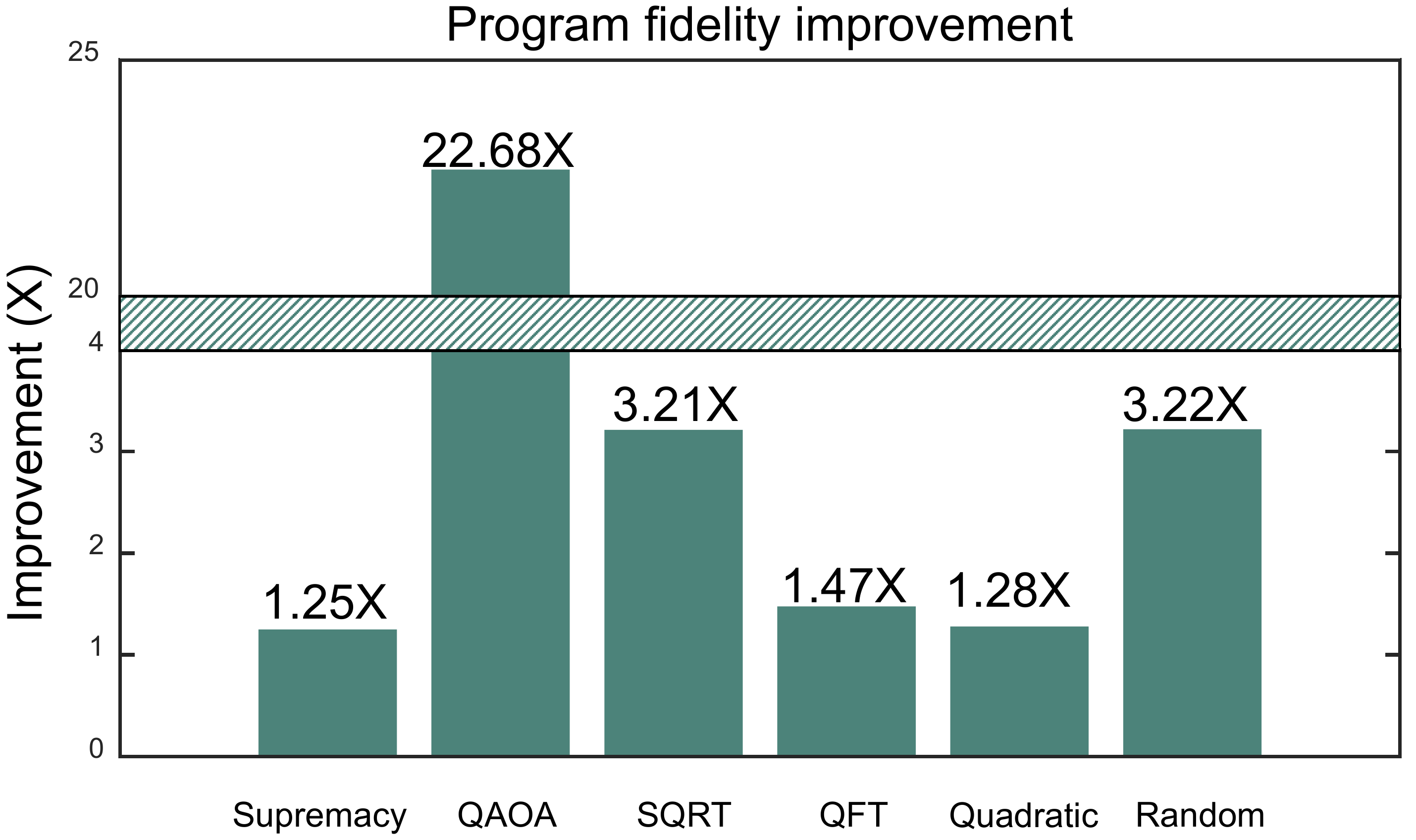}
    \caption{Improvement in program fidelity compared to~\cite{murali-ti}.\vspace{-11mm}}
    \label{fig:time-fid}
\end{figure}

Shuttle operation increases vibrational energy or motional mode($\bar{n}$) of an ion-chain i.e., heats up ion-chain and degrades gate fidelity. 
As our compiler optimizations reduce shuttles, it curbs motional mode resulting in improved gate and overall program fidelity. Fig.~\ref{fig:time-fid} shows improvement in program fidelity across different benchmarks. QAOA exhibits a very high-gain as it requires the highest number of shuttles across benchmarks. In general, applications with high shuttle-to-gate ratio will experience more improvement in program fidelity. This is because the fidelity of such applications are dominated by shuttle-induced aggravated motional mode (last term of the fidelity equation, Section~\ref{sec:basics}) than the background heating. However, reducing shuttle operations still significantly improves program fidelity for benchmarks with low shuttle-to-gate ratio, e.g, QFT.

\subsection{Compilation time overhead}
Our compiler optimizations for shuttle direction, gate reordering, and efficient re-balancing increase the compilation time compared to~\cite{murali-ti}. 
Our algorithms have worst case time complexity of $O(n^2)$. However, as we discuss in Section \ref{subsubsec:complexity-shuttle-dir}, \ref{subsubsec:complexity-reorder}, and \ref{subsubsec:complexity-nnf}, the value of $n$ is contained. Therefore, the compilation time remains tractable even for very large circuits like QFT and QuadraticForm (3000-4000 gates) as evident from the Table.~\ref{tab:comp-time-overhead}. 
For all the circuits the increase in compilation times are in few tens of seconds, and it remains under a minute for very large circuits. Therefore, we are trading-off compilation time in a scalable manner to reduce \# operations.

\begin{table}[t]
\centering
\caption{Compilation time overhead}
\label{tab:comp-time-overhead}
\begin{minipage}{8cm}
\centering
\def\arraystretch{1.5}\tabcolsep 2pt
\def\thefootnote{a}\footnotesize

\begin{tabular}{cccc}
\hline
Benchmark & Compile time (sec) & Compile time (sec) & $\Delta (\uparrow)$\\
 & [This work] & \cite{murali-ti} & (sec)\\
\hline
Supremacy   & 2.6 & 1.1 & 1.5\\
QAOA        & 12.99 & 3.88 & 9.11\\
SquareRoot  & 6.29 & 1.83 & 4.46\\
QFT         & 18.42 & 4.22 & 14.2\\
QuadraticForm       & 24.55 & 3.74 & 20.81\\
Random          & 19.15 (12.59) & 3.53 & 15.62 (11.28)\\

\hline
\vspace{-6mm}
\end{tabular}
\end{minipage}
\end{table}

\subsection{Discussions}

\subsubsection{Heuristic vs. exact methods}
Applying exact methods like integer linear programming (ILP) and satisfiability modulo theorem (SMT) solvers can lead to best results. However, these methods do not scale well with circuit size. For the NISQ-era benchmarks considered in this paper or the previous paper~\cite{murali-ti} the exact approaches will become intractable. Therefore, majority proposals on quantum compilers (e.g.,~\cite{qiskit-transpiler, gushu-li}) resort to heuristics methods and trades off some performance in the favor of better scalability. We also follow the same approach.

\subsubsection{Test benchmark sizes}
We test benchmarks with $60-75$ qubits and thousands of 2-qubit gates. One can argue that these benchmarks are not of practical sizes for present-day noisy devices. However, the domain of quantum computing is in a trajectory where benchmarks of this size will become and must become practical. Our compiler generates more compact circuits with substantially less number of operations (shuttles) which will definitely be beneficial.

\subsubsection{Initial mapping policy}
In this paper, we used popular greedy initial mapping policy~\cite{murali-noise-adaptive}. In future, different initial mapping policies can be explored. Even without modifying the initial mapping policy, our compiler shows significant gains.

\section{Conclusion}\label{sec:conclusion}
In this paper, we present compiler optimizations for multi-trap TI quantum computers. Our methods drastically reduce the number of shuttles compared to previous state-of-the-art and improves program fidelity.

\bibliographystyle{IEEEtran}
\bibliography{IEEEabrv, ref}

\end{document}